\input phyzzx
\hfuzz 20pt

\font\mybb=msbm10 at 12pt
\def\bb#1{\hbox{\mybb#1}}

\def\bR {\bb{R}}
\def\bE {\bb{E}}

\def\bH {\bb{H}}

\def\bfomeg{\omega\kern-7.0pt\omega}
\def\bfOmeg{\Omega\kern-8.0pt\Omega}

\REF\belavin{ A.A. Belavin, A.M. Polyakov, 
A.S. Schwartz \& Yu.S. Tyupkin, {\sl
Pseudoparticle Solutions of the Yang-Mills Equations}, 
 Phys. Lett. {\bf B59} (1975) 85.}
\REF\hooft{G. 'Hooft, {\sl Computation of the
Quantum Effects Due to the
Four-Dimensional Pseudoparticle}, Phys. Rev. 
{\bf D14} (1976) 3432.} 
\REF\adhm{M.F. Atiyah, V.G. Drinfeld, N.J. Hitchin \& Yu.I. Manin, 
{\sl Construction of
Instantons}, Phys. Lett. {\bf A65} (1978) 185.}
\REF\nuyts{ E. Corrigan, C. Devchand, D.B. Fairlie \& J. Nuyts, 
{\sl First-Order
Equations for Gauge Fields in Spaces of Dimension Greater than Four}, 
Nucl. Phys. {\bf
B214} (1983) 452.}
\REF\ward{R.S. Ward, {\sl Completely Solvable Gauge-Field
 Equations in Dimension
Greater than Four}, Nucl. Phys. {\bf B236} (1984) 381.}
\REF\kent{E. Corrigan, P. Goddard \& A. Kent, {\sl Some Comments 
on the ADHM Construction
in 4k Dimensions}, Commun. Math. Phys. {\bf 100} (1985) 1.}
\REF\PT{G. Papadopoulos and P.K. Townsend, {\sl Intersecting M-branes}, 
Phys.
Lett. {\bf 380B} (1996) 273.}
\REF\eric{E. Bergshoeff, M. de Roo, E. Eyras,
 B. Janssen \& J.P. van der
Schaar, {\sl Multiple Intersections of D-branes 
and M-branes}, hep-th/9612095.}
\REF\pope{H. Lu, C.N. Pope, T.A. Tran \& K.W. Xu, 
{\sl Classification of p-branes,
Nuts, Waves and Intersections}, hep-th/9708055.}
\REF\BDL{M. Berkooz, M.R. Douglas \& R.G. Leigh, 
{\it Branes Intersecting at
Angles}, Nucl. Phys.  {\bf B480} (1996) 265.}
\REF\ptg{J.P. Gauntlett, G.W. Gibbons, G. Papadopoulos 
\& P.K. 	Townsend, {\sl
Hyper-K\" ahler manifolds and multiply intersecting branes}, 
Nucl. Phys. (1997), to
appear: hep-th/9702202.}
\REF\cvetic { K. Behrndt \& M. Cveti\v c, 
{\sl BPS-Saturated Bound States of Tilted
p-Branes in Type II String Theory}, Phys. Rev. 
{\bf D56} (1997) 1188.}
\REF\myers{J.C. Breckenridge, G. Michaud and
 R.C. Myers, {\sl New Angles on Branes},
hep-th/9703041.}
\REF\costa{ M. Costa \& M. Cveti\v c, 
{\sl Non-Threshold D-Brane Bound States and
Black Holes with Non-Zero Entropy}, hep-th/9703204.}
\REF\larsen{V. Balasubramanian, F. Larsen \& R.G. Leigh, 
{\sl Branes at Angles and
Black Holes}, hep-th/9704143.}
\REF\myersb{G. Michaud and R.C. Myers, {\sl Hermitian 
D-Brane Solutions},
hep-th/9705079.}
\REF\banks{T. Banks, W. Fischer, S.H. Shenker and L. 
Susskind, {\sl M Theory as
a Matrix Model}, hep-th/9610043.}
\REF\motl {L. Motl,~~ {\sl Proposals on Nonperturbative
 Superstring Interactions},~~
hep-th/9701025.}
\REF\banksseib{T. Banks and N. Seiberg, {\sl Strings 
from Matrices}, hep-th/9702187.}
\REF\verlinde{R. Dijkgraaf, E. Verlinde and H. Verlinde, 
{\sl Matrix String Theory},
hep-th/9703030.}
\REF\hulltown{C.M. Hull and P.K. Townsend, {\sl Unity of
 Superstring Dualities}, Nucl.
Phys. {\bf B438} (1995) 109.}
\REF\ori{O.J. Ganor, S. Ramgoolam and W. Taylor IV, {\sl Branes, 
Fluxes and Duality in
M(atrix)-Theory}, hep-th/9611202.}
\REF\banksb{T. Banks, N. Seiberg and S. Shenker, 
{\sl Branes from Matrices},
hep-th/9612157.}
\REF\hulla{S.J. Gates, C.M. Hull \& M. Ro\v cek, 
{\sl Twisted Multiplets and New
Supersymmetric Nonlinear Sigma Models}, Nucl. Phys. 
{\bf B248} (1984) 157.}
\REF\HP {P.S. Howe \& G. Papadopoulos, {\sl Ultraviolet
 Behaviour of Two-Dimensional
Nonlinear Sigma Models}, Nucl. Phys. {\bf B289} (1987) 264; 
{\sl Further Remarks on the
Geometry of Two-Dimensional Nonlinear Sigma Models}, Class. 
Quantum Grav. {\bf 5} (1988)
1647.}
\REF\howepapd{P.S. Howe \& G. Papadopoulos, 
{ \sl Twistor Spaces for HKT manifolds},
Phys. Lett.  {\bf B379} (1996) 80.}
\REF\lambert{N.D. Lambert, {\sl Heterotic p-Branes 
from Massive Sigma Models}, Nucl.
Phys. {\bf B477} (1996) 141, hep-th/9605010.}
\REF\donaldson {S.K. Donaldson, {\sl Anti Self-dual 
Yang-Mills Connections over Complex
Algebraic Surfaces and Stable Vector Bundles}, Proc. 
London Math. Soc. {\bf 50} (1985) 1.}
\REF\nair {V.P. Nair \& J. Schiff, {\sl A K\"ahler-Chern-Simons 
Theory and
Quantisation of Instanton Moduli Spaces}, Phys. Lett. {\bf 246B} 
(1990) 423.} 
\REF\rothe{J.J. Giambiaggi \& K.D. Rothe, {\sl Regular
N-Instanton Fields and Singular
Gauge Transformations}, Nucl. Phys. {\bf B129} (1977) 111.}
\REF\gps {G.W. Gibbons, G. Papadopoulos \& K.S. Stelle, 
{\sl HKT and OKT Geometries
on Black Hole Soliton Moduli Spaces}, hep-th/9706207.} 
\REF\howepapb{P.S. Howe \& G. Papadopoulos, {\sl Finiteness 
and Anomalies in (4,0)
Supersymmetric Sigma Models},  Nucl .Phys. 
{\bf B381} (1992) 360. }
\REF\hermannone{S. Fubini \& H. Nicolai, {\sl The 
Octonionic Instanton}, Phys. Lett.
{\bf 155B} (1985) 369.}
\REF\ivanova{ T. A Ivanova, {\sl Octonions, Self-Duality 
and Strings}, Phys. Lett. {\bf
B315} (1993) 277.}
\REF\hermanntwo{M. G\"unaydin \& H. Nicolai, {\sl Seven-Dimensional 
Octonionic Yang-Mills
Instanton and its Extension to an Heterotic String Soliton}, 
Phys. Lett. {\bf B351}
(1995) 169.}
\REF\zhou{ N. Ohta and J-G. Zhou, {\sl Realization of D4-Branes at Angles
in Super Yang-Mills Theory}, hep-th/9709065.}


\Pubnum{ \vbox{ \hbox{DAMTP-R/97/43} } }
\date{August 1997}
\pubtype{Revised, November 1997}

\titlepage

\title {\bf Instantons at Angles }

\author{G. Papadopoulos and A. Teschendorff}
\address{DAMTP, University of Cambridge,
\break
Silver Street,
\break
Cambridge CB3 9EW, U.K.}

\abstract{We interpret a class of 4k-dimensional instanton 
solutions found by Ward,
Corrigan, Goddard and Kent as 
four-dimensional instantons at
angles. The superposition of each pair of four-dimensional
instantons is associated with four angles which  
depend on some of the ADHM
parameters.  All these solutions
 are associated with
the group
$Sp(k)$ and are examples of Hermitian-Einstein 
connections on
$\bE^{4k}$. We show that the eight-dimensional solutions preserve $3/16$
of the ten-dimensional N=1 supersymmetry. 
We  argue that under the correspondence between the BPS states of Yang-Mills
theory 
and those of M-theory that arises in the context of Matrix models, 
the instantons
at angles configuration  corresponds to the longitudinal intersecting 5-branes
on a string at angles configuration of M-theory.  }

\endpage
\pagenumber=2
\sequentialequations

\chapter{Introduction}

Instantons are Yang-Mills configurations
 which are characterized by the first Pontryagin number, 
$\nu$, and obey 
the antiself-duality condition\foot{We have chosen the 
antiself-duality condition in
the definition of the instanton but our results can be easily
 adapted to the
self-duality one.}
$\star F=-F$, 
where  $F$ is the field strength of a gauge potential $A$, the  
star is the Hodge star
and $\star
\star=1$ for Euclidean signature 4-manifolds.
One key property of this condition is that together
with the Bianchi identities, $\nabla_{[M} F_{NL]}=0$, 
it {\sl implies} the field
equations of the Yang-Mills theory.   At first,  a large class
 of instanton solutions was
found  using  the  ansatz of [\belavin,
\hooft]. Later a systematic  classification 
of the solutions of the
antiself-duality conditions  was done in [\adhm], 
which has become known as the ADHM
construction. 
Subsequently, many generalizations of  the self-duality condition beyond
four dimensions have been proposed and for some of 
them ADHM-like constructions have
been found [\nuyts, \ward, \kent].

The understanding of non-perturbative
aspects of various superstring theories involves the
investigation of configurations which have the interpretation
of intersecting branes. In the context of the effective
theory, these are extreme
solutions of various supergravity theories [\PT].
 Such solutions are constructed
using powerful superposition rules from the \lq elementary' 
 brane solutions
that preserve $1/2$ of the spacetime supersymmetry
 (for recent work see [\eric,
\pope] and references within).  One feature of the  
intersecting branes  is
that the intersections can occur at angles in 
such a way that the
configurations preserve a proportion of supersymmetry 
[\BDL]. Solutions of various
supergravity theories that have the interpretation of
 intersecting branes at angles have
been found in [\ptg-\myersb].

More recently, a non-perturbative formulation of 
M-theory and string theory was proposed using the 
Matrix models [\banks, \motl , \banksseib, \verlinde]. These involve the 
study of certain
Yang-Mills theories on a torus which is dual to the compactification
 torus of 
M-theory. The U-duality groups [\hulltown]
 of IIA superstring theory
 are naturally
realised in the context of matrix models [\ori].
As a result, there is a correspondence between the BPS
states of these Yang-Mills theories [\banksb] and those of
M-theory and superstring theories. In the low energy 
effective theory, the BPS states 
of the latter can
be described by various classical solitonic solutions of various
supergravity theories.  This indicates that there is a
correspondence between BPS states of Yang-Mills theories
with the solitonic solutions of supergravity theories.
Since in some cases, the solitonic BPS states of Yang-Mills theories
can be described by classical solutions that carry the appropriate
charges, the above argument suggests that there is a correspondence between
solitonic solutions of supergravity theory with solitonic solutions
of Yang-Mills theory. Although this correspondence
is established for a compactification torus of finite size,
we expect that the correspondence between the solutions of the
Yang-Mills theory and  supergravity theory will persist in the various limits
where the torus becomes very large or very small. One
example of such correspondence is the following: The 
longitudinal five-brane of the
Matrix model is described 
 either as the five-brane solution
of D=11 supergravity
superposed with a pp-wave and compactified on a four-torus or as an instanton
 solution of the Yang-Mills theory (see for example
 [\ori, \banksb]) on the dual four-torus. (The pp-wave carries
 the longitudinal
momentum of the five-brane.) Ignoring the size of the 
compactifying torus, one
can simply say that there is a correspondence between 
longitudinal five-branes and
instantons.  One purpose of this letter
 is to demonstrate the
correspondence between BPS solutions of the Yang-Mills
theories with BPS solutions of the supergravity theories
with another example.
In D=11 supergravity theory, there is a configuration with
the interpretation of intersecting five-branes at
 angles [\ptg] on a string. This
configuration is associated with the group $Sp(2)$; 
the proportion of
supersymmetry preserved  is directly related
to the number of singlets in the decomposition of the spinor representation 
in eleven dimensions under $Sp(2)$. Superposing
this configuration with a pp-wave along the
 string directions corresponds
to a Matrix theory configuration with the
interpretation of {\sl intersecting longitudinal
fivebranes at angles}. Now using the correspondence
between the longitudinal fivebranes and the Yang-Mills instantons
mentioned above, one concludes that there should {\sl exist}
solutions of Yang-Mills theory that have the interpretation
of {\sl instantons at angles}.  Moreover since the instanton configuration
must preserve the same proportion of supersymmetry as the supergravity
 one, it must
also be associated with the group $Sp(2)$.

To superpose Yang-Mills 
 BPS configurations, like for example
Yang-Mills instantons or monopoles, at angles, we embed them as 
solutions of an 
appropriate higher-dimensional
Yang-Mills theory.  Such solutions will be localized 
on a subspace of the associated
spacetime. Then given such a configuration we shall
 superpose it with another similar 
one which, however, may be localized in a different subspace. We  require
 that the solutions of the
Yang-Mills equations which have the interpretation of
 BPS configurations at angles to have 
the following properties: (i) They solve a BPS-like  
equation  which reduces to the
standard BPS conditions of the original BPS configurations
 in the appropriate dimension 
(ii) if the solution which describes the superposition
 can be embedded in a
supersymmetric theory, then it preserves a proportion
 of supersymmetry. 

In this letter,  we shall describe the superposition of 
four-dimensional instantons. Since instantons are associated
 with four-planes, it is
clear that the configuration that describes the superposition
 of two instantons at an
angle is eight-dimensional; in general the configuration  for k
instantons lying on linearly independent four-planes is 4k-dimensional. 
It turns out that
the appropriate BPS condition in $4k$ dimensions
 necessary for the superposition
of four-dimensional instantons has being found by Ward
[\ward]. Here we shall describe how this BPS condition is 
associated with the group
$Sp(k)$.  This allows us to interpret a class of the 
$4k$-dimensional instanton solutions
of Corrigan, Goddard and Kent [\kent] as 
four-dimensional instantons at angles. We shall find that there 
{\sl four angles} associated
with the superposition of each pair of two 
four-dimensional instantons which 
depend on the ADHM
parameters. The eight-dimensional solutions preserve
 $3/16$ of the N=1 ten-dimensional
supersymmetry and correspond to the  intersecting longitudinal
fivebranes at angles configuration of the
 eleven-dimensional supergravity.

\chapter{Tri-Hermitian-Einstein Connections}

The BPS condition of Ward [\ward] is naturally associated with the
group $Sp(k)$. For this, we first observe that the curvature 
$$
F_{MN}=\partial_MA_N-\partial_NA_M+ [A_M,A_N]\ ,
\eqn\inthree
$$
of a connection $A$ on $\bE^{4k}$ can be thought
as an element of $\Lambda^2(\bE^{4k})$ with respect to
the space indices. Now $\Lambda^2(\bE^{4k})=so(4k)$ where
$so(4k)$ is the Lie algebra of $SO(4k)$.  Decomposing $so(4k)$ 
under $Sp(k)\cdot Sp(1)$, we find that
$$
\Lambda^2(\bE^{4k})=sp(k)\oplus sp(1)\oplus \lambda^2_0\otimes \sigma^2
\eqn\decompot
$$
where we have set $\Lambda^1(\bE^{4k})=\lambda\otimes \sigma$ 
under the decomposition
of one forms under $Sp(k)\cdot Sp(1)$, $\lambda^2_0$ 
is the symplectic traceless
two-fold anti-symmetric irreducible representation
 of $Sp(k)$ and $\sigma^2$ is the
symmetric two-fold irreducible representation of $Sp(1)$.
The integrability condition of [\ward] can be 
summarized by saying that the
components of the curvature along the last two
 subspaces in the decomposition
\decompot\ vanish.  This implies that the space
indices of the curvature
take values in $sp(k)$.  It is convenient to
 write this BPS condition
in a different way.  Let $\{I_r\, ; r=1,2,3.\}$ 
be three complex structures
in $\bE^{4k}$ that obey the algebra of imaginary
 unit quaternions, {\sl i.e.}
$$
I_r I_s=-\delta_{rs}+\epsilon_{rst} I_t\ .
\eqn\inseven
$$ 
The BPS condition then becomes
$$
(I_r){}^P{}_M (I_r){}^R{}_N F_{PR}{}^a{}_b=F_{MN}{}^a{}_b\ ,
\eqn\gsd
$$
(no summation over $r$), {\sl i.e.} the curvature 
is (1,1) with respect
to all complex structures. We shall refer to 
the connections that satisfy
this condition as tri-Hermitian-Einstein connections.
 We remark that this is
precisely the condition required by (4,0) supersymmetry 
 on the
 curvature of connection of the Yang-Mills sector
in two-dimensional sigma models\foot{For the application of the
instantons of [\kent] in sigma models see 
[\lambert].} [\hulla, \HP, \howepapd].

It is straightforward to show that \gsd\ reduces
 to the
 antiself-duality condition in four dimensions\foot{In fact 
it reduces to either
the self-duality or the antiself-duality condition depending
 on the choice of
orientation of the four-manifold.} and that it {\sl implies} the 
Yang-Mills field equations in 4k
dimensions. Moreover any solution of \gsd\ is an example of
an Hermitian-Einstein (or Hermitian Yang-Mills) connection\foot{For $k>1$, 
the Hermitian-Einstein BPS
condition is weaker than that of \gsd.} (see for
 example [\donaldson, \nair]). The
latter are connections for which (i) the curvature
 tensor is (1,1) with respect to a 
complex structure and (ii)
$$ 
g^{\alpha\bar\beta} F_{\alpha\bar\beta}=0\ ,
\eqn\trac
$$
where $g$ is a hermitian metric. 
We remark that the above  two conditions  for 
Hermitian-Einstein
 connections on
$\bR^{2n}$ imply that the non-vanishing components
 of the curvature are in 
$su(n)$. (The condition \trac\ together 
with the Bianchi identity imply
the Yang-Mills field equations.)  The 
connections that satisfy \gsd\ are 
Hermitian-Einstein with respect to three 
different complex structures. 

\chapter{One-angle Solutions}

To interpret some of the solutions of [\ward] 
and [\kent] as four-dimensional
instantons at angles, we first use an appropriate
generalization of the t'Hooft ansatz. For this, we write
$\bE^{4k}=\bE^4\otimes \bE^k$, {\sl i.e.} we 
choose the coordinates
$$
\{y^M\}=\{x^{i\mu};\, i=1,\dots,k ;\, \mu=0,1,2,3\}\ ,
\eqn\inten
$$ 
on $\bE^{4k}$.  Then we introduce three complex structures on $\bE^{4k}$ 
$$
({\bf I}_r){}^{i\mu}{}_{j\nu}= I_r{}^\mu{}_\nu \delta^i{}_j
\eqn\inonea
$$
where $\{I_r;\, r=1,2,3\}$ are  three complex structures on 
$\bE^4$ associated with the
K\"ahler forms
$$
\eqalign{
\omega_1&=dx^0\wedge dx^1+dx^2\wedge dx^3
\cr
\omega_2&=-dx^0\wedge dx^2+dx^1\wedge dx^3
\cr
\omega_3&=dx^0\wedge dx^3+dx^1\wedge dx^2\ ,}
\eqn\kahler
$$
respectively. Choosing the volume form as
 $\epsilon=dx^0\wedge dx^1\wedge dx^2\wedge
dx^3$, these form a basis of self-dual 2-forms 
in $\bE^4$. We also write the
Euclidean metric on $\bE^{4k}$ as
$$
ds^2=\delta_{\mu\nu} \delta_{ij} dx^{i\mu} dx^{j\nu}\ .
\eqn\intwoa
$$
Similarly, the curvature two-form is
$$
F_{MN}=F_{i\mu j\nu}\ .
\eqn\inthreea
$$
Writing
$$
F_{i\mu j\nu}=F_{[\mu\nu]}{}_{(ij)}+ F_{(\mu\nu)}{}_{[ij]}\ ,
\eqn\inout
$$
we find that \gsd\ implies that  
$$
\eqalign{
{1\over2}F_{[\mu\nu]}{}_{(ij)}
\epsilon^{\mu\nu}{}_{\rho\sigma}&=-F_{[\rho\sigma]}{}_{(ij)}
\cr
F_{(\mu\nu)}{}_{[ij]}&=\delta_{\mu\nu} f_{ij} \ ,}
\eqn\inin
$$
where $f_{ij}$ is a $k\times k$ antisymmetric matrix.
 Next, we shall further assume that the gauge
group is
$SU(2)=Sp(1)$ and write the ansatz
$$
A_{i\mu}=i \Sigma_\mu{}^\nu \partial_{i\nu} {\rm log}\phi(x)
\eqn\ansatz
$$
where the matrix two-form $\Sigma_{\mu\nu}$ is the
 'Hooft tensor, {\sl
i.e.}
$$
\Sigma_{\mu\nu}={1\over2}\omega_{r\mu\nu} \sigma^r\ ,
\eqn\infoura
$$
and $\{\sigma_r; r=1,2,3\}$ are the Pauli matrices.
 After some computation, 
we find that the field
strength
$F$ of 
$A$ in
\ansatz\ satisfies the first equation in \inin\ provided that
$$
{1\over \phi}\partial_i\cdot\partial_j \phi=0\ .
\eqn\condition
$$
This is a harmonic-like equation and a solution is
$$
\phi=1+{\rho^2\over (p_i\, x^{i\mu}-a^\mu)^2}\ ,
\eqn\onepi
$$
where $\{p\}=(p_1,p_2, \dots, p_k)$ are $k$ real numbers, 
$a^\mu$ is the centre of the
harmonic function and $\rho^2$ is the parameter associated
 with the size of the
instanton. More general solutions can be obtained by a 
linear superposition of the above
solution for different choices of $\{p\}$ and $a^\mu$ leading to
$$
\phi=1+\sum_{\{p\}} \sum_{ a} {\rho^2_a(\{p\})\over \big(p_i\, 
x^{i\mu}-a^\mu(\{p\})\big)^2}\ .
\eqn\solution
$$
Finally, it is straightforward to verify that the 
curvature of the connection \ansatz\
 with $\phi$ given as in  \solution, satisfies \gsd. 

To investigate the properties of the above solutions, 
let us suppose that the harmonic
function
$\phi$ involves only one
$\{p\}$ as in
\onepi. In this case it is always possible, after 
 a coordinate transformation,
to set
$\{p\}=\{1, 0,
\dots, 0\}$.  Then
$\phi$ becomes a harmonic function on the \lq first' copy 
of $\bE^4$ in $\bE^{4k}$.  The
resulting configuration is that derived from the ansatz
 of [\belavin, \hooft] on 
$\bE^4$. In particular, the instanton number is equal
 to the number of centres
$\{a^\mu\}$ of the harmonic function. In this case the
 singularities at the
 centres of the harmonic
function can be removed with a
 suitable choice of a gauge
transformation [\rothe]. Next,
let us suppose that our solution  involves two
 linearly independent vectors
$\{p\}$, say $\{p\}=\{p\}$ and $\{p\}=\{q\}$. 
If we take $|p_i x^i|\rightarrow
\infty$, then the configuration reduces to the 
solution associated with a harmonic
function that involves only $\{q\}$.  As we have explained above, 
this is the instanton
solution of  [\belavin, \hooft] on an appropriate 4-plane in $\bE^{4k}$. 
Similarly if
we take  $|q_i x^i|\rightarrow
\infty$, then the configuration reduces to the solution 
associated with the harmonic
function that involves only $\{p\}$. We can therefore 
 interpret this solution as
the superposition of two four-dimensional instantons at angles 
in $\bE^{4k}$.  The angle is   
$$
\cos\theta={p\cdot q\over |p| |q|}\ .
\eqn\ineighta
$$
 As it may have been expected from
 the non-linearity of the gauge potentials in 
\gsd, this superposition is non-linear even in the case for which $\{p\}$
and $\{q\}$ are orthogonal.
The solution appears to be singular at the positions of the
harmonic function.  For simplicity let us consider two 
instantons at angles with
instanton number one. In this case we have two centres.
 The singular set is defined by
the two codimension-four-planes
$$
\eqalign{
p_i\, x^{i\mu}-a^\mu&=0
\cr
q_i\, x^{i\mu}-b^\mu&=0\ ,}
\eqn\inhun
$$
in $\bE^8$.
We have investigated the singularity structure of the
 solution when $\{p\}$ and $\{q\}$
are orthogonal using a method similar to [\rothe].  We
 have found that
these singularities can be removed everywhere apart from 
the intersection of the two
singular sets. A similar observation
has been made in [\kent].  To conclude, let us briefly consider the 
  general case.  The number
of four dimensional instantons involved in the configuration
 is equal to the number of
linearly independent choices for
$\{p\}$. Their instanton number is equal to the number of
 centres associated with each
choice of $\{p\}$. The general solution describes the superposition of
 these four-dimensional
instantons at  angles.

\chapter{Four-angle solutions}

The ADHM ansatz of [\kent] describes solutions
 with more parameters than those found in
the previous section using the t'Hooft ansatz. 
Here we shall interpret some of these
parameters as angles. In fact we shall find that 
there are {\sl four angles} 
between each
pair of planes associated with these 
solutions all determined in terms of ADHM
parameters. For simplicity, we shall take $Sp(1)$ as 
the gauge group. The ADHM ansatz is
$$
A=v^\dagger d v\ ,
\eqn\adhmconn
$$
where $v$ is an $(\ell+1)\times 1$ matrix of 
quaternions normalized as
$v^\dagger v=1$,
and $\ell$ is an integer which is identified with 
the `total' instanton number (see
[\kent]). The matrix
$v$ satisfies the condition
$$
v^\dagger \Delta\equiv v^\dagger (a+b_i x^i)=0
\eqn\adhmcond
$$
where  $a, b_i$ are $(\ell+1)\times \ell$ matrices 
of quaternions;  we have arranged the 
4k coordinates into quaternions $\{x^i\, ;
i=1,\cdots, k\}$ ($\bE^{4k}=\bH^k$). The connection \adhmconn\ satisfies
the  condition \gsd\
 provided that the matrices
$a^\dagger a, b_i^\dagger b_j, a^\dagger b_i$ are
 symmetric as matrices of quaternions. 

A convenient choice of matrices $a, b_i$ leads to
$$
\eqalign{
(\Delta)_{0n}&=-a_0 \lambda_n
\cr
(\Delta)_{nm}&=(p^\dagger_{ni} x^i-a_n) \delta_{nm}}
\eqn\newone
$$
where $\{\lambda_n; n=1,\dots, \ell\}$ are real numbers, and
$\{p_{ni}; n=1,\dots, \ell; i=1,\dots, k\}$  and $\{a_0, a_n;
n=1,\dots,\ell\}$ are quaternions . The solutions
 discussed in the previous
section  using the t'Hooft ansatz correspond to
 choosing $\{p_{ni}; n=1,\dots, \ell,
i=1,\dots, k\}$
to be real numbers.  For $v$, we
find 
$$
\eqalign{
(v)_0&=-{a_0\over |a_0|}  f^{-1}
\cr
(v)_n&=- |a_0|\, \lambda_n\, \big[(x^i)^\dagger\, 
p_{ni} -a^\dagger_n\big]^{-1} \,
f^{-1}}
\eqn\newtwo
$$
where  $f$ is the normalization
factor 
$$
f^2=1+  |a_0|^2\, \sum^\ell_{n=1}  {\lambda_n^2\over
 |p^\dagger_{ni} x^i-a_n|^2}\ .
\eqn\newthree
 $$
We remark that the sum in the normalization factor
 is over the centres $\{a_n;
n=1,\dots, \ell\}$.
It is clear from the form of the normalization
 factor that this solution is associated
with $\ell$ codimension-four-planes in $\bE^{4k}$.  
The equations of these planes are
$$
p^\dagger_{ni}\, x^i-a_n=0\ ,
\eqn\newfour
$$
for $n=1, \dots, \ell$.  The four normalized
 normal vectors to these planes  are
$$
{\cal N}_n={1\over |p_n|}\, 
(p_n^i)^\dagger{\partial\over \partial x^{i}}\ ,
\eqn\normalone
$$
in quaternionic notation, where 
$|p_n|^2= \delta_{ij}(p_n^i)^\dagger\, p_n^j$.
Next let us consider two such
 codimension-four-plane determined say by $p=p_1$ and
$q=p_2$ and with associated normal 
vectors ${\cal N}$ and $\tilde {\cal N}$,
respectively. The angles associated with 
these planes are given by the inner
product of their normal vectors, so
$$
\cos(\theta)=(\tilde {\cal N})^\dagger\, {\cal N}=
{\delta_{ij} (q^i)^\dagger\, p^i\over |p|
|q|}\ ,
\eqn\moreangles
$$ 
where $\cos(\theta)$ is a quaternion in the obvious notation. 
 There are four angles
because  $\cos(\theta)$ has four components. 
These four angles are all different for a generic
 choice of $p,q$ and independent from
the choice of gauge fixing for the residual symmetries of 
the ADHM construction  [\kent].
It is worth mentioning that each pair of
 four-dimensional instantons
are superposed at $Sp(2)$ angles. For this 
we observe that the normal vectors
\normalone\ of the two codimension-four-planes
 span an eight-dimensional subspace
in $\bE^{4k}$ and their coefficients are quaternions. 
So, they can be related with an
$Sp(2)$ rotation (after choosing a suitable
 basis in $\bE^{4k}$). As a result the two
four-planes spanned by the normal vectors in
$\bE^8$ are at
$Sp(2)$ angles. Such four-planes are parameterized
 by the four-dimensional coset space
$Sp(2)/SO(4)$;  this explains the presence of 
four angles in \moreangles.
It also suggests that there may be a 
generalization of the solutions of [\ptg] to
describe the superposition five-branes 
and KK-monpoles  depending on four angles. 

A naive counting of the dimension of 
the moduli of the 4k-dimensional solutions
 which takes account of  the dimension of moduli of
4-dimensional instantons involved in the 
superposition and their associated angles
{\sl does not} reproduce the dimension
 of the moduli space in [\kent].
However, all their solutions have a decaying 
behaviour at large distances consistent
with the interpretation that they are 
four-dimensional instantons at angles.

\chapter{Supersymmetry}

Among the above Yang-Mills configurations 
on $\bE^{4k}$ only the instantons at angles in
 $\bE^8$ and the instantons on $\bE^4$ 
are solutions of 
ten-dimensional supersymmetric  Yang-Mills theory  preserving a
proportion of  supersymmetry. 
The supersymmetry condition is
$$
F_{MN} \Gamma^{MN}\epsilon=0\ ,
\eqn\intena
$$
where $\epsilon$ is the supersymmetry parameter. 
In the four-dimensional
case the instantons preserve $1/2$ of the supersymmetry.
In the eight-dimensional case condition 
\gsd\ implies that 
$F_{MN}$ is in $sp(2)$,    an argument
similar 
to that in [\ptg] can be used to show that the instantons 
 at  angles in $\bE^8$ will preserve $3/16$ of 
supersymmetry.  As a solution of the Matrix theory,  
the eight-dimensional
instantons at angles preserve $3/32$ of 
spacetime supersymmetry. This
is in agreement with the proportion of 
supersymmetry preserved by the
intersecting five-branes on a string at 
angles and superposed with a pp-wave solution
of D=11 supergravity. However there is a difference. In 
the case of two orthogonal
intersecting five-branes on a string with a pp-wave 
superposed the proportion of the
supersymmetry preserved is
$1/8$ but this is not the case for the configuration of 
two orthogonal instantons. To see this, first 
observe that the  solutions of the
previous two sections
will preserve $1/4$ of the N=1 ten-dimensional
 supersymmetry if the components of
the curvature are in the $sp(1)\oplus sp(1)$ 
subalgebra of $sp(2)$. However this is not
so because a direct calculation reveals that 
the mixed components
$F_{\mu1,\nu2}$ of the curvature tensor 
do not vanish.   Further we remark that one might
have thought that it is possible to superpose 
two four-dimensional instantons
$B_\mu$ and
$C_\alpha$ with gauge group $G$ localised at 
two orthogonal four-planes in $\bE^8$ by
simply setting
$A=(B,C)$ and requiring that the  gauge group 
of the new connection is again $G$. 
However, this is not a solution of the BPS condition \gsd\ for a 
non-abelian gauge group in
eight-dimensions. This is because  \gsd\ is
non-linear in the connection. However the above
linear superposition is a solution if the gauge groups of $B$ and $C$ 
are treated
independently, i.e. if the  gauge group of $A$ is $G\times G$.

\chapter{Concluding Remarks}

We have interpreted some of the 4k-dimensional instantons
of [\ward] and [\kent] as  superposition of four-dimensional
instantons at angles.  We have shown that
 there {\sl four angles} associated with the
superposition of two four-dimensional 
instantons which depend on the ADHM parameters.
These solutions
 are examples of
Hermitian-Einstein connections in 4k
 dimensions and the eight-dimensional ones
preserve $3/16$ of the N=1 ten-dimensional
 supersymmetry.  Because of the close relation
between the self-duality condition and  the BPS equations for
magnetic  monopoles we expect that one can generalize the 
above construction to find
solutions of Yang-Mills equations on $\bE^{3k}$ which have
 the interpretation of
monopoles at angles.  In addition, the Yang-Mills configurations
of [\kent] and the  HKT geometries described in [\gps] combined give
new examples of two-dimensional
supersymmetric sigma models with (4,0) supersymmetry. These 
provide consistent backgrounds for the propagation of the heterotic
string [\howepapb]. Another application of the instantons of [\kent] is in 
the context of D-branes.  In IIA theory, 
they are associated with the D-brane bound
state of a 0-brane within a 4-brane 
within an 8-brane, and in the IIB theory they are
associated with the D-brane bound
 state of a D-string within a D-5-brane within a
D-9-brane.

There are many other ways to superpose two or more four-dimensional
instantons.  Here we have described the
 superposition using the condition that the components of
the curvature $F$ of the Yang-Mills connection are in  $sp(k)$.
Another option is to use the Hermitian-Einstein 
condition associated with $su(2k)$ which we have
already mentioned in section two.  In
 eight dimensions this will lead to configurations
preserving $1/8$ of the N=1 
ten-dimensional supersymmetry. In addition in eight dimensions one can
allow the components of $F$ to lie in  $spin(7)$.  This will result in
 superpositions of
instantons preserving $1/16$ of the N=1 ten-dimensional
 supersymmetry.  In fact, some 
$spin(7)$ instanton solutions have already been
 found [\hermannone]. It will be of
interest to see whether they can be interpreted 
as four-dimensional instantons at
angles. Since $Sp(2)$ is a subgroup of $Spin(7)$,
 the solutions of [\kent] are also
examples of $spin(7)$ instantons, 
albeit of a particular type. There
are more possibilities in lower dimensions,
 for example the
$G_2$ instantons in seven dimensions [\ivanova, \hermanntwo]
 which may have a similar
interpretation.

\vskip 1cm

\leftline{NOTE ADDED}

While we were revising our work, the paper [\zhou] by N. Ohta and J-G Zhou
 appeared 
in which  the  relation between supergravity 
configurations and Yang-Mills ones
is also discussed. However, they use constant Yang-Mills
 configurations
 on the eight
torus.

\vskip 1cm
\noindent{\bf Acknowledgments:}   We would like to thank
 P. Goddard and A. Kent for
helpful discussions. A.T. thanks PPARC for a 
studentship. G.P. is supported by a
University Research Fellowship from the Royal Society.

\refout
\bye